\documentclass[twocolumn]{aastex6}
\usepackage{amsmath}
\usepackage{xcolor}
\usepackage{natbib}
\usepackage{rotating}
\usepackage{graphicx}                           
\usepackage{amssymb}

\begin{document}

\title{Rare Event Statistics Applied to Fast Radio Bursts}
\author{Scott Vander Wiel} 
\affil{Statistical Sciences, Los Alamos National Laboratory, Los Alamos, NM}
\author{Sarah Burke-Spolaor}
\affil{National Radio Astronomy Observatory, Socorro, NM}
\author{Earl Lawrence} 
\affil{Statistical Sciences, Los Alamos National Laboratory, Los Alamos, NM}
\author{Casey J. Law} 
\affil{Department of Astronomy and Radio Astronomy Lab, University of California, Berkeley CA}
\and
\author{Geoffrey C. Bower}
\affil{Academia Sinica Institute of Astronomy and Astrophysics, 645 N. A'ohoku Place, Hilo, HI 96720, USA}

\begin{abstract}
Statistical interpretation of sparsely sampled event rates has become vital for new transient surveys, particularly those aimed at detecting fast radio bursts (FRBs). We provide an accessible reference for a number of simple, but critical, statistical questions relevant for current transient and FRB research and utilizing the negative binomial model for counts in which the count rate parameter is uncertain or randomly biased from one study to the next.  We apply these methods to re-assess and update results from previous FRB surveys, finding as follows.
1)~Thirteen FRBs detected across five high-Galactic-latitude ($>30^\circ$) surveys are highly significant ($p=5\times10^{-5}$) evidence of a higher rate relative to the single FRB detected across four low-latitude ($<5^\circ$) surveys, even after accounting for effects that dampen Galactic plane sensitivity.  High- vs.~mid-latitude ($5^\circ$ to $15^\circ$) is marginally significant ($p=0.03$).
2)~A meta analysis of twelve heterogeneous surveys gives an FRB rate of 2866 sky$^{-1}$day$^{-1}$ above 1~Jy at high Galactic latitude (95\% confidence 1121 to 7328) and 285 sky$^{-1}$day$^{-1}$ at low/mid latitudes (95\% from 48 to 1701).
3)~Using the Parkes HTRU high-latitude setup requires 193 observing hours to achieve 50\% probability of detecting an FRB and 937 hours to achieve 95\% probability, based on the ten detections of \citep{champion2016fivenew} and appropriately accounting for uncertainty in the unknown Poisson rate.
4)~Two quick detections at Parkes from a small number of high-latitude fields \citep{ravi2015fast, petroff2015survey} tentatively favor a `look long' survey style relative to the `scan wide' HTRU survey, but only at $p=0.07$ significance.
\end{abstract}

\maketitle

\section{Introduction}
Current interest in transients surveys, including those for Fast Radio Bursts (FRBs), often begs for estimation and comparison of rare event rates using multiple surveys with small numbers of detections, including zero detections \citep{thornton2013population,tingay2015mwafrbs,champion2016fivenew,stewart2016claus}. Publications and proposals make statistical confidence statements about event rates, probabilities of detecting events, and inconsistancies of rates across different studies. This  paper fills the need for an easily accessible reference that recommends and illustrates statistical methods for estimating and comparing rates from Poisson counts of rare events.  Uncertainty or variation in the underlying event rates is conveniently represented through negative binomial models for count data.  We discuss methods for answering the following questions:
\begin{itemize}
	\item How many observing hours are needed for a high probability of observing a pulse with the Parkes multibeam receiver?
	\item Do surveys with a few longer pointings produce FRBs at a higher rate?
	\item Is the rate of FRBs smaller in low or mid Galactic latitudes compared to high latitudes?
	\item Do multiple FRB surveys collectively follow a common underlying rate?  What rate?
	\item What is the trend of FRB rate with flux limit?
\end{itemize}
We recommend both Bayesian and frequentist methods with the goal of keeping the analysis solid and simple.  A convenient Bayesian formulation is used to answer the first three questions easily.  For the next two questions Bayesian solutions would require Monte Carlo Markov chain simulations, so we switch to frequentist methods that are readily available in  statistical analysis software.

Our exposition is framed mostly in terms of FRB transients but the methods are more generally applicable to rare transient events.  The key assumption is that  events follow a Poisson process, meaning that the total number of events in the observed sky$\times$time domain is a Poisson random variable and the sky$\times$time coordinates of multiple events are independent and identically distributed.
Mostly we assume the homogeneous case of uniformly distributed coordinates in which each infinitesimal sky$\times$time volume element contributes an event with probability proportional to its size. The Poisson process assumption requires that events must be rare relative to their spatiotemporal extent so that probabilities of overlapping events can be ignored.  
Because the rate function of the Poisson process is unknown and possibly observed with survey-specific biases, we use extensions of the simple Poisson distribution to estimate underlying Poisson rates.

This paper follows the standard practice of utilizing count statistics across a number of surveys to estimate FRB rates and illustrates the use of either simple calculations or a widely available generalized linear model setup to produce valid rate estimates.  Our companion paper, \citet{lawrence2016}, introduces a new method that utilizes the same count data to estimate a rate trend with respect to sensitivity limit but also incorporates observed flux values for each of the detections.  That new method requires special purpose statistical model fitting but it has several advantages too: valuable flux data contribute to estimates of the rate, the source count (power-law) index is better constrained, and beam shapes are explicitly incorporated into the fitting rather than being approximated as top hat functions.  Some detailed comparisons are provided in Section~\ref{sec:NBRegression}. 

This paper is organized as follows.   Section~\ref{sec:data} provides key parameters of thirteen FRB surveys, six of which did not discover an FRB.  Section~\ref{sec:hours} introduces the negative binomial model for a Poisson count with uncertain rate and applies it to calculate  FRB discovery time at Parkes.  Section~\ref{sec:twoGroups} extends the statistical setup to describe uncertainty in the ratio of event rates for two groups of studies and applies this to compare rates in surveys with long vs.~short pointings and low, mid, and high Galactic latitudes.  Section~\ref{sec:extraPoisson} uses the negative binomial model to represent cross-survey counts that are more varied than would be expected from a single common (but unknown) Poisson rate.  A likelihood ratio hypothesis test, applied to the data of Table~\ref{tab:data}, shows that common rates are plausible for separate high- and combined low/mid-Galactic-latitude analyses, but not for the fully combined set under an assumption of Euclidian scaling.  Section~\ref{sec:rateChange} illustrates the use of negative binomial regression to estimate a rate function (vs.~flux sensitivity) with an estimated source count index, Galactic latitude dependence, and possible extra-Poisson variation.  The regression estimate is broadly consistent with rates reported by other authors.  Finally, Section~\ref{sec:summary} summarizes the methods and conclusions.

\section{Survey Data Used in This Paper}\label{sec:data}

Table~\ref{tab:data} lists data from thirteen papers that report on FRB searches; this is the same set of surveys analyzed by~\citet{lawrence2016}.  
\defcitealias{burke2014galactic}{Burke-Spolaor~et~al.~(2014)}


\begin{sidewaystable*}[pht]
\centering \vspace*{3in}
\begin{tabular}{rlllrrrrrrr}
Label & Reference\footnote{Data listed for Lor07, Sie12, Tho13, Spi14, Bur14, Pet15, Rav15, and Cha16 are consistent (within rounding) with Table~2 of \citet{burke2016limits}, except for Tho13 in which time is taken from the original reference and exposure is rounded to the value cited in \citet{champion2016fivenew}.  Data for Kea10, Bur12, Pet14, and Den19 are taken from Table~2 of \citet{burke2014galactic} with sensitivity limits computed in the same manner as \citet{burke2016limits}.  Observation time for \citet{petroff2014absence} is $1157-231$ hr, where the reduction by 231~hr is for pointings not sensitive to FRBs, following the authors.  Data for Law15 are taken directly from the reference.}  & Facility & Galactic\footnote{Galactic Latitude is a coarse categorization of the survey's pointings: low, $|b|\leq 5^\circ$; mid, $5^\circ<|b|<15^\circ$; high, $b\geq 30^\circ$.}  & $N_\text{FRB}$ & Sensitivity\footnote{Sensitivity limit is  at half the maximum beam gain, double the beam-center values given by~\citet{lawrence2016}.} & Diameter\footnote{Diameter is the FWHM value.} & Beams & Time & Exposure & $E_\text{1Jy}$\footnote{$E_\text{1Jy}$ is the equivalent exposure at 1~Jy sensitivity under Euclidian scaling, Equation~(\ref{eq:euclidian}).} \\
   &  &  & Latitude& [count] & [Jy] & [arcmin] & [count] & [hr] & [deg$^2$ hr] & [deg$^2$ hr] \\
 \hline
Lor07 & \citet{lorimer2007bright} & Parkes & high &  1 & 0.590 & 14 & 13 & 490.5 & 272.7 & 601.7 \\ 
  Den09 & \citet{deneva2009arecibo} & Arecibo & low &  0 & 0.365 & 4 & 7 & 459.1 & 8.6 & 39.0 \\ 
  Kea10 & \citet{keane2010further} & Parkes & low &  0 & 0.447 & 14 & 13 & 1557.5 & 865.8 & 2897.1 \\ 
  Bur12 & \citet{burgay2012perseus} & Parkes & low &  0 & 0.529 & 14 & 13 & 532.6 & 296.1 & 769.5 \\ 
  Sie12 & \citet{siemion2012search} & ATA & all &  0 & 118.000 & 150 & 30 & 135.8 & 19998.2 & 15.6 \\ 
  Bur14 & \citetalias{burke2014galactic} & Parkes & mid &  1 & 0.615 & 14 & 13 & 917.3 & 509.9 & 1057.3 \\ 
  Pet14 & \citet{petroff2014absence} & Parkes & mid &  0 & 0.868 & 14 & 13 & 926.0 & 514.8 & 636.5 \\ 
  Spi14 & \citet{spitler2014fast} & Arecibo & low &  1 & 0.210 & 4 & 7 & 11503.3 & 215.2 & 2236.2 \\ 
  Law15 & \citet{law2015millisecond} & VLA & high &  0 & 0.240 & 30 & 1 & 166.0 & 32.6 & 277.2 \\ 
  Pet15 & \citet{petroff2015survey} & Parkes & high &  1 & 0.555 & 14 & 13 & 85.5 & 47.5 & 114.9 \\ 
  Rav15 & \citet{ravi2015fast} & Parkes & high &  1 & 0.555 & 14 & 13 & 68.4 & 38.0 & 91.9 \\ 
  Cha16 & \citet{champion2016fivenew} & Parkes & high & 10 & 0.560 & 14 & 13 & 2786.5 & 1549.0 & 3696.3 \\ 
   \hline
Tho13 & \citet{thornton2013population} & Parkes & high &  4 & 0.560 & 14 & 13 & 551.9 & 316.0 & 754.1 \\ 

  \end{tabular}

\caption{Exposure and detection data for thirteen FRB surveys with 15 total detections.  Tho13 is set apart because it is subsumed by Cha16.}
\label{tab:data}
\end{sidewaystable*}

 The surveys were conducted on four telescopes and collectively detected 15 FRBs, with 10 of these coming from the high Galactic latitude portion of the High Time Resolution Universe (HTRU) survey at Parkes \citep{petroff2015survey}.  \citet{thornton2013population}, listed last, is contained within \citet{champion2016fivenew}.  V-FASTR results \citep{burke2016limits} are not considered here because they are reported with an analysis method that could not be readily reduced to the summary form of Table~\ref{tab:data}. Subsequent sections use portions of these data to address questions of interest in current FRB research and illustrate statistical methods.   The table provides a short label and citation, the facility used for observation, Galactic latitude category, number of FRBs detected, flux sensitivity (at the half-max beam radius; see notes on fluxes in next paragraph), beam diameter (FWHM), number of beams, time on sky (per beam), exposure (area$\times$time), and $E_\text{1Jy}$, the equivalent exposure at 1 Jy sensitivity under Euclidian scaling, as discussed in Subsection~\ref{sec:anisotropic}.   The categorization of Galactic latitudes is coarse, indicating the range of the large majority of pointings in each study: low ($|b|\leq 5^\circ$), intermediate ($5^\circ<|b|<15^\circ$), or high ($b\geq 30^\circ$).
 
Flux sensitivities listed in Table~\ref{tab:data} are determined from the calculation methods of \citet{burke2014galactic} and \citet{burke2016limits}. Those works include sensitivity corrections that account for losses due to different instrumental configurations between surveys including the noise contribution of the position-dependent Galactic sky temperature. They also incorporate the typical scattering and dispersion that would influence an FRB in a survey of the given sky coverage, i.e.\ FRBs discovered in Galactic plane surveys will on average have greater levels of scattering and, particularly, dispersion. Those corrections are reflected in these numbers.

\section{How many hours are needed to observe a pulse?}\label{sec:hours}
In this section we show how to determine the number of hours required to detect an event with given probability, while accounting for uncertainty in the event rate.

For comparison we first illustrate a method that is incomplete but typically used. As an example,\citet{thornton2013population} discovered four FRB pulses in the high-Galactic latitude region from initial observations of the High Time Resolution Universe (HTRU) survey \citep{keith2010high} using the Parkes multibeam receiver.  
\citet{champion2016fivenew} re-processed these observations with an upgraded pipeline, finding Thornton's four FRBs and one more; they also processed further HTRU high-latitude observations finding five additional pulses, for a total of ten FRB detections.

\citet{thornton2013population} reported four pulses in 551.9 hours of observation. 
A common practice is to assume a known underlying rate of $R=4/551.9=0.00725$ pulses per hour and use Poisson probabilities for detecting $x$ pulses in $T$ hours of observation,
\begin{equation}
	\Pr(x) = \frac{(RT)^x e^{-RT}}{x!}, 
\end{equation}
 to claim that $T=413$ hours will give a 95\% chance of success because
\begin{align}\label{eq:atLeastOnePoisson}
	\Pr(&\text{at least one pulse}) \nonumber \\
	& = 1-\Pr(x=0) = 1-e^{-0.00725\cdot 413} = 0.95.
\end{align}	
But 413 hr is too low because it assumes no uncertainty in $R$, which is obviously a stretch when the rate was estimated from only four previous detections.  For reference, the time for a 50\% chance is 95.6 hr under this Poisson assumption. 

The following Bayesian setup admits uncertainty in $R$ and produces a direct requirement on observing time with a modest burden to specify an appropriate prior distribution for the unknown rate.  Subsection~\ref{sec:ParkesTime} answers the Parkes question.

\subsection{Negative Binomial Model for a Poisson Count with an Uncertain Rate}\label{sec:negbin}
Designate past and future observation periods with respective subscripts $i=1,2$ and let
\begin{subequations}\label{eq:notation}
\begin{align}
        X_i =& \text{ count of events,} \\
        T_i =& \text{ observation time [hr],} \\
 	A_i =& \text{ effective beam area [deg$^2$],} \\
        S_i =& \text{ product of known sensitivity factors such as } \nonumber \\
                 & \text{ processing efficiency, $(\text{flux sensitivity})^{-3/2}$, } \nonumber \\
                 & \text{ wavelength effect},
                 \label{eq:sensitivity} \\
	E_i =&\ T_i A_i S_i = \text{exposure  [deg$^2$ hr]}, \label{eq:exposure}
\end{align}
\end{subequations}
and
\begin{align}
        R =& \text{ unknown true event rate [deg$^{-2}$hr$^{-1}$] }  \nonumber \\
             & \text{ under unit sensitivity ($S=1$).}
\end{align}
Each sensitivity factor comprising $S_i$ should represent a multiplicative effect on the rate of events.  Furthermore, the flux sensitivity (i.e.~minimum detectable flux) might, itself, incorporate loss and efficiency factors, such as in \citet[Section~4]{rane2016search}.

The goal is to determine $T_2$ to achieve a conditional probability
\begin{equation}
	\Pr(X_2 \geq 1 \ |\ X_1) = 0.95.
\end{equation}
For a given rate $R$, model the counts as 
\begin{align}\label{eq:xi}
        [X_i |R] &\ \stackrel{\text{ind}}{\sim}\  \text{Poisson($RE_i$)} \quad(i=1,2).
\end{align}
The probability notation says that, for given $R$, event counts $X_1$ and $X_2$ are independent Poisson random variables with expectations $RE_i$. 
We describe {\em a priori} uncertainty in the unknown rate $R$ with the gamma distribution: 
\begin{align}
        R &\sim \text{Gamma}(\alpha,\beta)
\end{align}
where the Gamma probability density function is parameterized as $\beta^\alpha R^{\alpha-1} e^{-\beta R} / \Gamma(\alpha)$ with $\Gamma(\cdot)$ denoting the gamma function.  A recent recommendation with good justification is to use values $\alpha=1/3$ and $\beta \rightarrow 0$ for a so-called {\em neutral} prior on $R$; this is especially apropos for small counts in situations where little or no outside information about $R$ is available; \citet{kerman2011neutral} provide details regarding this choice of values, and the implications of using other values.  

The Bayesian analysis turns out to be simple because the gamma prior is  {\em conjugate} to the Poisson likelihood, meaning the posterior distribution of $R$ remains in the gamma family.  Standard  results \citep[Section~2.7]{gelman2014bayesian} \citep[Chapter~3]{hoff2009first} determine posterior distributions for $R$ and $X_2$, conditioned on $X_1$:
\begin{align}
        \label{eq:RgivenX}
        [R  | X_1] &\ \sim\ \text{Gamma}(\alpha+X_1, \beta+E_1)  \text{\ (conjugate)}\\ 
        \label{eq:X2givenX1}
        [X_2  | X_1]  &\ \sim\ \text{NB}\left(r=\alpha+X_1, \ p=\frac{\beta+E_1}{\beta+E_1+E_2}\right)
\end{align}
with negative binomial (NB) probabilities parameterized as
\begin{align}\label{eq:NB}
	\Pr(x;r,p) &= \frac{\Gamma(r+x)}{\Gamma(x+1)\Gamma(r)} p^r (1-p)^x
\end{align}
for $(x=0,1,2,\ldots)$.  Therefore, 
\begin{align}
        \Pr(&\text{at least one new event}) \nonumber \\
        &= 1 - \Pr(X_2=0 \ |\ X_1) 
         = 1-p^r \nonumber \\
        & = 1- \left(\frac{\beta+T_1 A_1 S_1}{\beta+T_1 A_1 S_1+T_2 A_2 S_2}\right)^{\alpha+X_1}.
        \label{eq:probOne}
\end{align}
The right-hand side of Equation~(\ref{eq:probOne}) can also be interpreted as a cumulative distribution function (CDF) for time, $T_2$, to detect an FRB.  This form of CDF is known as the Pareto Type~II distribution. 

\subsection{Observing Time at Parkes}\label{sec:ParkesTime}

So, how much Parkes multibeam time is required to discover an FRB?  The answer evolves  as more studies are published and processing algorithms are improved.  We give two answers: one based on only the \citet{thornton2013population} study and a second using the complete \citet{champion2016fivenew} data.

First, assume the four \citet{thornton2013population} detections and set $X_1=4, T_1=551.9, A_1 S_1=A_2 S_2, \alpha=1/3$, and $\beta=0$ to get
\begin{align}\label{eq:atLeastOne}
        \Pr(&\text{at least one new event}) \nonumber \\
        &= 1 - \left(\frac{551.9}{551.9+T_2}\right)^{1/3+4}.
\end{align}
A 95\% chance of success is obtained by plugging in $T_2=550$ hours, 33\% higher than the 413 hours based on the Poisson probabilities of Equation~\ref{eq:atLeastOnePoisson}.  Recognition that $R$ is uncertain requires more time for high confidence relative to an assumed known rate.  On the other hand, the NB 50\% detection time is 95.7 hr, essentially equal to the Poisson calculation following Equation~(\ref{eq:atLeastOnePoisson}), so the 50\% time is hardly affected by rate uncertainty.

With the full HTRU high-latitude results of \citet{champion2016fivenew} (10 FRBs in 2786.5 hours), the  time required is now 937 hr for 95\% probability of detection using NB calculations, 12\% more than 835 hr obtained using a Poisson calculation.  
The current 50\% detection time is 193 hr, identical within rounding using either Poisson or NB calculations.

\begin{figure}[tbp]
\begin{center}	
	{\center 
	\includegraphics[width=\columnwidth]{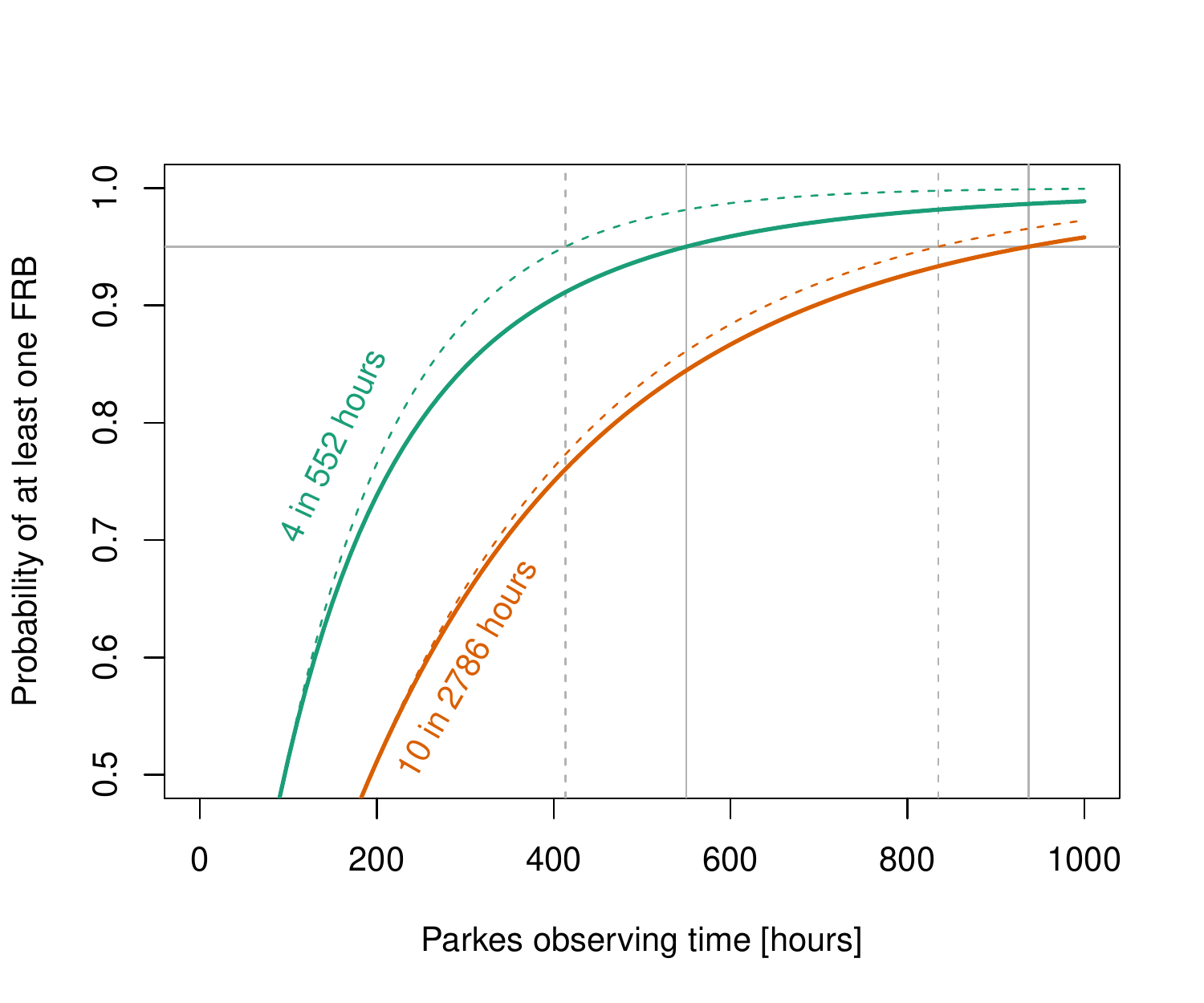}}
	\caption{Probability of detecting at least one FRB verses observing time, as assessed after four (cyan) and then ten (orange) detections.   Solid curves are negative binomials that reflect uncertainty in the underlying rate whereas dashed curves are Poissons that assume exactly known rates equal to observed values.}
	\label{fig:parkes}
\end{center}
\end{figure}

Figure~\ref{fig:parkes} expands this, plotting the probability to detect at least one FRB as a function of time on sky as assessed after the four detections of \citet{thornton2013population} (cyan) and then after processing the further observations of \citet{champion2016fivenew} (orange).  Solid curves are based on negative binomials with a functional form as exemplified in Equation~(\ref{eq:atLeastOne}); these account for uncertainty in the rate.  Dashed curves assume known rates and straight Poisson calculations like Equation~(\ref{eq:atLeastOnePoisson}).  Grey vertical lines indicate the number of observing hours to achieve a probability of 0.95, as in the example calculations above.  The initial Thornton set delivered a higher rate than the later full set of high-latitude HTRU data, so the full-data curves shift to the right, even as differences between Poisson and NB are reduced.  As time on sky accumulates the true rate is better known and eventually the Poisson calculation becomes adequate.  

The Appendix illustrates a frequentist method of determining time to detection that demands almost double (!) the hours for 95\% confidence, using the Thornton data, relative to the Bayesian calculation. We argue (in the appendix) that this is too conservative.  On the other hand, note that  the Bayesian answer did, in fact, double once the full HTRU observations were incorporated.  In subsequent sections we use a combination of Bayesian and frequentist methods with the choice giving heavy consideration to simplicity.

\section{Is One Style of Survey More Productive Than Another? } \label{sec:twoGroups}

This section addresses questions comparing two groups of surveys: (1) Do surveys with longer pointings produce FRBs at a higher rate?; and (2) Is the rate of FRBs larger in high Galactic latitudes compared to low and mid latitudes? The answers are `yes' and `yes' with marginal statistical significance for (1) and high significance for (2), as shown in Subsections~\ref{sec:wideDeep} and~\ref{sec:anisotropic}.
Differences among surveys could arise from a number of factors, including routine statistical variation,  actual differences in underlying rates of transients, or some unidentified observational bias separating the groups.  Next, we show how to compare rates with a formal statistical analysis.  

Extend the notation in Equation~(\ref{eq:notation}) to include include two sets of surveys and model their event counts $\mathbf{X} \equiv (X_1,\ldots, X_K)$ and $\mathbf{Y} \equiv (Y_1, \ldots, Y_L$) as
\begin{subequations}
\begin{align}
        [X_i |R_x] &\ \stackrel{\text{ind}}{\sim}\  \text{Poisson($R_xE_{xi}$)} \quad (i=1,\ldots,K) \\
        [Y_i |R_y] &\ \stackrel{\text{ind}}{\sim}\  \text{Poisson($R_yE_{yi}$)} \quad (i=1,\ldots,L) \\
        R_x, R_y  &\ \stackrel{\text{ind}}{\sim}\  \text{Gamma}(\alpha,\beta).
\end{align}
\end{subequations}
Sensitivity factors can be multiplied into the exposures as described in Equation~(\ref{eq:sensitivity}) to account for known differences between the groups of surveys.  The following analysis determines if the underlying rates differ even after accounting for such known effects.

Equation~(\ref{eq:RgivenX}) extends to provide  (statistically independent) posterior distributions of $R_x$ and $R_y$ 
\begin{subequations} \label{eq:twoRates}
\begin{align}
        [R_x \ |\ \mathbf{X}, \mathbf{Y}] &\sim   \Gamma(\alpha+X_\cdot, \beta+E_{x\cdot}) \label{eq:Rx}, \\
        [R_y \ |\ \mathbf{X}, \mathbf{Y}] &\sim   \Gamma(\alpha+Y_\cdot, \beta+E_{y\cdot}) \label{eq:Ry}
\end{align}
\end{subequations}
where $X_\cdot = X_1+\cdots+X_K$ and $Y_\cdot = Y_1+\cdots+Y_L$ are the total numbers of events and $E_{x\cdot}$  and $E_{y\cdot}$ are the corresponding total exposures.  

Our interest is to compare  rates $R_x$ and $R_y$.  A standard result in probability states that the ratio of independent gamma distributions is a scaled F distribution.  In particular, the posterior distribution for the ratio of rates is \citep{price2000estimating}
\begin{equation} \label{eq:rateRatio}
 \left[\left. \frac{R_x}{R_y} \ \right|\ \mathbf{X}, \mathbf{Y}\right] \sim \frac{\hat R_x}{\hat R_y} \cdot {\cal F}(m=2\alpha+2X_\cdot, n=2\alpha+2Y_\cdot)
\end{equation}
where $\hat R_x \equiv (\alpha + X_\cdot)(\beta+E_{x\cdot})^{-1}$, $\hat R_y \equiv (\alpha + Y_\cdot)(\beta+E_{y\cdot})^{-1}$, and ${\cal F}(m,n)$ denotes an $F$-distributed random variable with $n$ numerator and $m$ denominator degrees of freedom.  Therefore a 95\% credible interval for the ratio of rates is 
\begin{equation} \label{eq:rateRatioInterval}
	\frac{\hat R_x}{\hat R_y} \cdot Q_F(0.025; n,m) \ \leq\  \frac{R_x}{R_y}\  \leq \ \frac{\hat R_x}{\hat R_y} \cdot Q_F(0.975;n,m).
\end{equation}
where $n$ and $m$ are indicated in Equation~(\ref{eq:rateRatio}) and $Q_F$ is the quantile function of the $F$ distribution.
If the interval contains the value 1.0, then it is plausible (at the 95\% level) that the two types of surveys have equal FRB detection rates. Finally, the posterior probability of the first survey style being less productive than the second is
\begin{align}\label{eq:rateRatioPval}
	\Pr[ R_x < R_y \ |\ \mathbf{X}, \mathbf{Y}] 
	& = \Pr\left[\frac{\hat R_x}{\hat R_y} \cdot{\cal F}(n,m) < 1 \right]   \nonumber \\
	& = P_F(\hat R_y/\hat R_x; n,m)
\end{align}
where $P_F$ is the $F$ cumulative distribution function.

This method assumes that common rates ($R_x,R_y$) {\em are} appropriate {\em within} each group of studies.  Section~\ref{sec:extraPoisson} explains how to test whether a common rate is plausible within a group.

\subsection{Is there a notable difference between `scan wide' and `look long' survey styles?}
\label{sec:wideDeep}

The HTRU high-Galactic latitude survey \citep{champion2016fivenew} consisted of more than 30k pointings, each 270 s in duration.  The survey discovered ten FRBs in 116 days of on-sky time---one every 11.6 days, on average.  On the same Parkes 13-beam instrument with a similar setup, \citet{ravi2015fast} found a single FRB in a short study with 2.8 days  of on-sky time by focusing on a single field around the Carina dwarf galaxy.  Similarly \citet{petroff2015survey} found one in a short study with 3.6  days of on-sky time by deep search of only 8 fields with previous FRB detections.   This raises the question: Is the `look long' style of observation notably more productive than the `scan wide' style?

\begin{figure}[tbp]
\begin{center}	
	\includegraphics[width=\columnwidth]{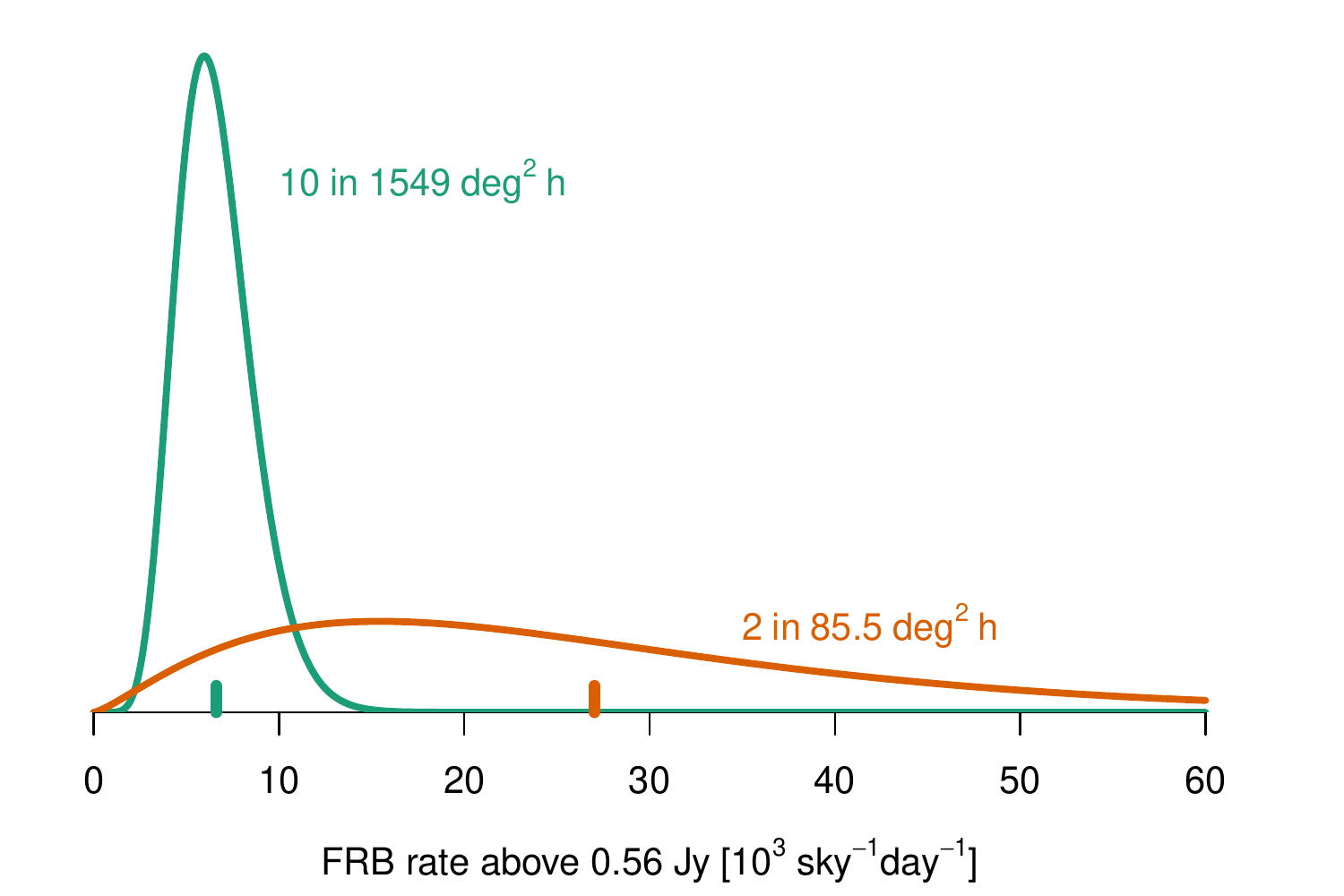} \\[-6ex]
	\includegraphics[width=\columnwidth]{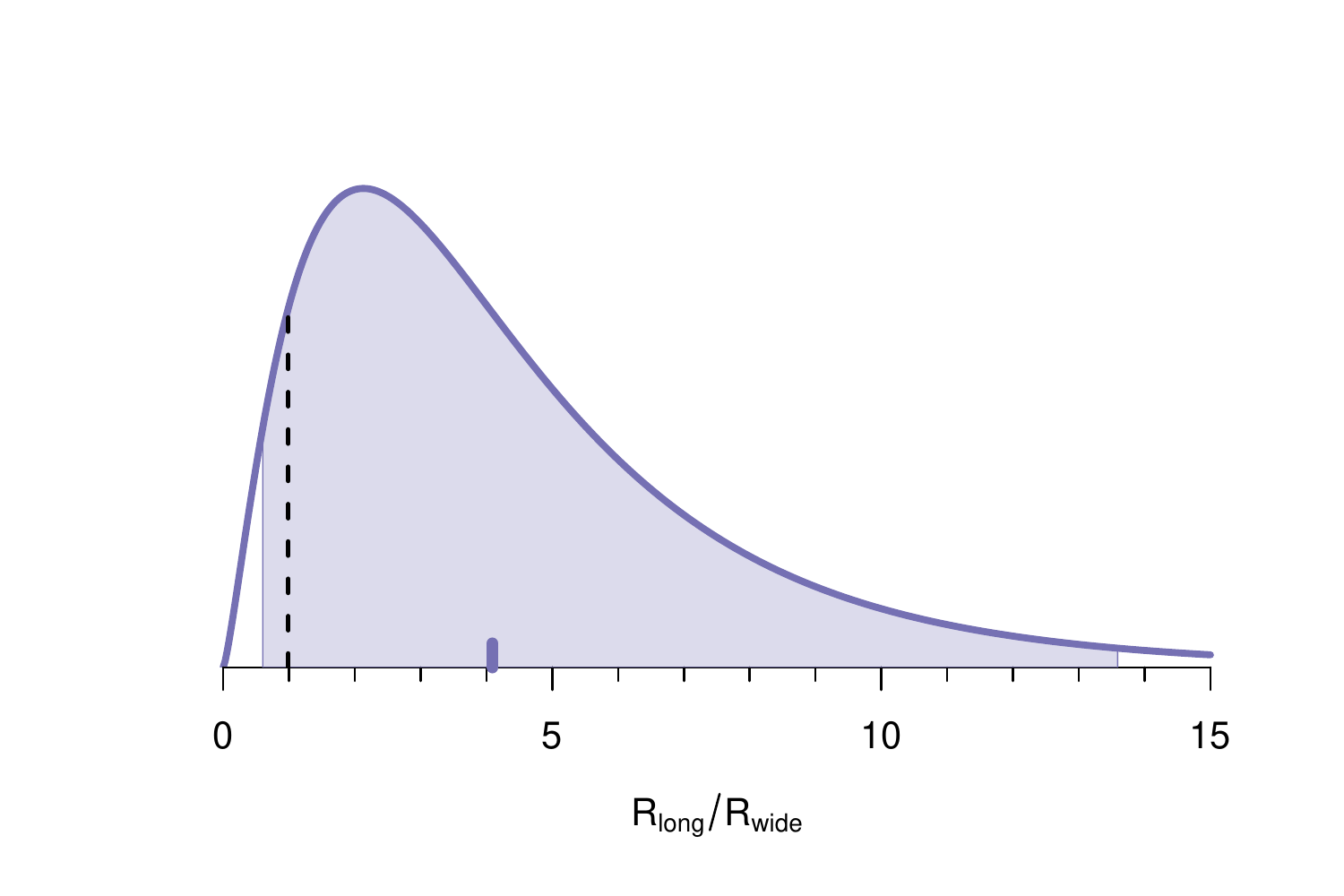} \\[-2ex]
	\caption{Top: Posterior probability density functions (pdfs) for the FRB rates in  a `scan wide' survey \citep[cyan, HTRU high latitude]{champion2016fivenew} and in the combination of two `look long' studies \citep[orange]{ravi2015fast, petroff2015survey}. The two `look long' detections suggest a higher rate but large uncertainty prevents a definite conclusion.  Bottom: Probability density function for the ratio of rates, $R_\text{long}/R_\text{wide}$.  The vertical line is at a ratio of one. Shading indicates 95\% credible bounds with the lower bound below a value of one.}
	\label{fig:wideDeep}
\end{center}
\end{figure}

Combining the two `look long' surveys, gives two detections from a cumulative exposure of $E_\text{long}=85.5$ deg$^2$~hr compared to ten detections from 
$E_\text{wide}=1,549$ deg$^2$~hr in the `scan wide' HTRU study \citep{champion2016fivenew}.   These surveys all have a flux sensitivity of 0.56 Jy.  The top panel of Figure~\ref{fig:wideDeep} shows the two posterior distributions of detection rates with the Parkes multi-beam setup, obtained by plugging exposures and FRB counts into Equation~(\ref{eq:twoRates}) with the neutral prior $(\alpha=1/3, \beta=0)$.  Colored marks on the bottom axis indicate means of the distributions. Although the nominal rates differ by a factor of 3.6, large uncertainty from the two `look long' surveys (orange) bridges across the narrower `scan wide' distribution (cyan).  Systematically different rates are not strongly indicated.

The bottom panel of Figure~\ref{fig:wideDeep} quantifies the result by plotting the probability density function for $R_\text{long}/R_\text{wide}$ using Equation~(\ref{eq:rateRatio}) with $(X,Y)=(2,10)$,  $(E_{x\cdot}, E_{y\cdot})= (E_\text{long},E_\text{wide})=(85.5, 1549)$.  The estimated ratio is  $\hat R_\text{long}/\hat R_\text{wide} = [(1/3+2)/85.5][(1/3+10)/1549]^{-1}= 4.1$ with 95\% credible interval (0.59, 13.6), calculated from Equation~(\ref{eq:rateRatioInterval}) and indicated by the shaded region.  The interval contains a ratio of 1.0 so the data suggest, but do  not demand, a conclusion of $R_\text{long} > R_\text{wide}$; Equation~(\ref{eq:rateRatioPval}) puts the posterior probability at 0.93 and we caution against coming up with hypotheses that match idiosyncrasies observed in a small number of surveys and retrofitting an astronomical theory.


\subsection{Are FRBs more rare in low and mid Galactic latitudes compared to high latitudes?}\label{sec:anisotropic}

\citet{petroff2014absence} concluded that the absence of FRB detections in the HTRU intermediate latitude survey was indicative of a significantly lower rate compared to the four high latitude detections reported in \citep{thornton2013population}  at the time.  \citet{burke2014galactic} arrive at a similar result in a joint analysis of multiple surveys at high, mid, and low  Galactic latitudes. Now, with further observations and employing our statistical framework, we ask: Are Galactic latitude difference still significant?   We compare the five high-Galactic-latitude surveys in Table~\ref{tab:data} to the two mid latitude and four low latitude surveys.  Recall that quoted sensitivity values have already taken into account the sensitivity loss implied for the additional sky temperature contributed by the Galaxy, and the added scattering and dispersion that is imparted to an extragalactic pulse by traveling through the Galaxy before being seen at Earth. These effects are more severe at lower Galactic latitudes, lowering their sensitivity to extragalactic pulses. As in \citet{burke2016limits}, these sensitivity corrections are based on the sky temperature model of \citet{Oliveira-Costaetal2008} and the Galactic electron density model of \citet{ne2001}.

If FRB pulses are
distributed uniformly through the Universe without significant luminosity evolution, then the rate of FRBs detectable above a given flux value $F$ is proportional to $F^{-3/2}$, where 3 is the volume increase with distance and 2 is the flux decay with distance.  This is known as Euclidian scaling. More generally, power-law scaling has the rate above $F$ proportional to $F^{-b}$ where $b>0$ is known as the source count index.

In this section, we assume Euclidian scaling and therefore utilize the 1 Jy equivalent exposures, $E_\text{1Jy}$, from Table~\ref{tab:data} computed as
\begin{equation}\label{eq:euclidian}
	E_\text{1Jy} = \text{Exposure} \left(\frac{\text{Sensitivity}}{1Jy} \right)^{-3/2}.
\end{equation}
The numbers of detections in high-, mid-, and low-latitude surveys are, respectively, (13, 1, 1) with cumulative (1 Jy) exposures of  
$(E_\text{high-lat}, E_\text{mid-lat}, E_\text{low-lat}) = (4782.0, 1693.8, 5941.8)$ deg$^2$~hr.

\begin{figure}[b]
\begin{center}	
	\includegraphics[width=\columnwidth]{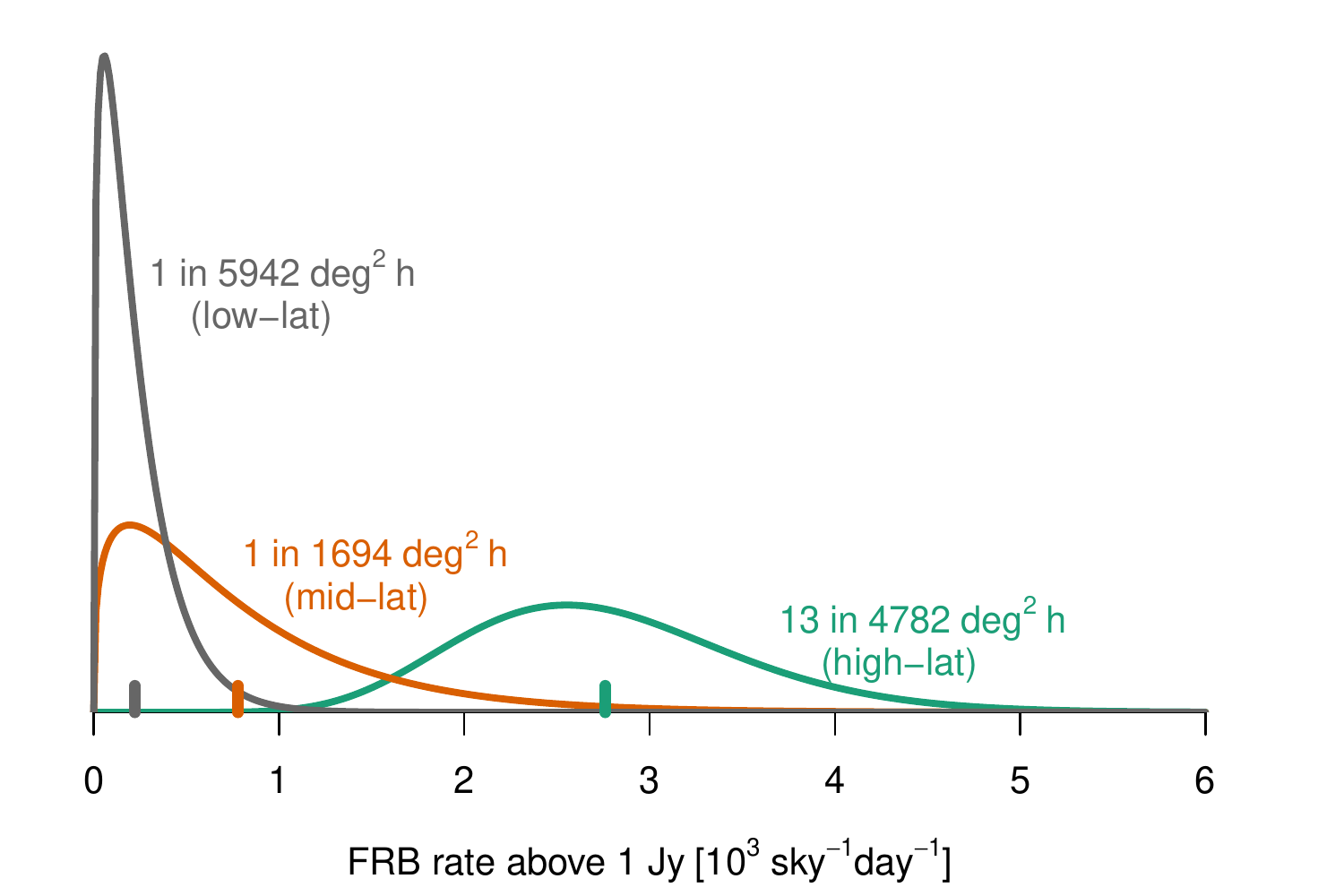}
\caption{Posterior probability density functions (pdfs) for the FRB rate in high- (cyan), intermediate- (orange), and low-Galactic latitude surveys (grey).   The low and high distributions are well-separated but mid overlaps both of the others.}
	\label{fig:highIntermediate}
\end{center}
\end{figure}

Figure~\ref{fig:highIntermediate} shows separate posterior distributions for the three rates using Equation~(\ref{eq:RgivenX}) with neutral priors $(\alpha=1/3, \beta=0)$.  The pdfs tell most of the story---namely, high- and low-lat are well-separated and mid-lat is closer to low but also overlaps high somewhat.  
Plugging $(X,Y)=(13,1)$ and $(E_{x\cdot},E_{y\cdot})=(4782.0,1693.8)$  into Equation~(\ref{eq:rateRatio}) produces the probability density function for $R_\text{mid}/R_\text{high}$ and Equation~(\ref{eq:rateRatioPval}) provides the posterior probability of $R_\text{mid} < R_\text{high}$ as $P_F(1/0.282; 2.67, 26.7)=0.968$.  Similarly $\Pr(R_\text{low} < R_\text{high})=0.99995$.  So, under the assumption of Euclidian scaling, we find very strong evidence of lower rates at lower Galactic latitudes.

Because the survey sensitivities used here have already been corrected for effects that dampen Galactic plane sensitivity, as previously reported in \citet{burke2014galactic}, this result is significant in that the \citet{ne2001} electron density model appears to not adequately explain the dependence of FRB rates on Galactic latitude. The remaining significant difference in rates at high and low latitudes may be accounted for by latitude-dependent scintillation \citep{MJ2015} and errors in the \citet{ne2001} model, among other possibilities.


\section{Does a collection of FRB surveys follow a common underlying rate?}\label{sec:extraPoisson}

Several papers \citep[e.g.,][]{burke2014galactic,rane2016search} have considered whether results across multiple surveys at multiple facilities are consistent with a single common rate, after accounting for known differences in sensitivities.  
Outside astronomy it is not unusual for rare event counts to vary substantially beyond what would be expected from strictly Poisson statistics and, in these cases, it is important to use a statistical model that admits random survey-to-survey biases.
Extra-Poisson variation is  known as {\em over-dispersion} in literature on generalized linear models (GLMs) including Poisson regression.  The statistical term over-dispersion is not related to frequency dispersion of electromagnetic waves; rather, it refers to additional variation beyond what is expected from a nominal (e.g., Poisson) statistical model.  
Here we describe how to check for such extra-Poisson variation.  The statistical procedure is described first and then applied to evaluate whether a common rate is indicated in subgroups of the surveys of Table~\ref{tab:data} under an assumption of Euclidian scaling.

The probabilistic setup supposes that each of $K$ surveys detects events at a distinct (unknown true) rate, $R_i$ ($i=1,\ldots,K$) that differs from the underlying cross-survey mean rate $R_0$.  Specifically,
\begin{align} 
	[X_i|R_i ] &\ \stackrel{\text{ind}}{\sim}\  \text{Poisson}(E_i R_i) \label{eq:XiGivenRi} \\
	R_i &\ \stackrel{\text{ind}}{\sim}\  \text{Gamma}\left(\alpha=r,\ \beta=r/R_0 \right) \label{eq:Ri} \\
	\Rightarrow \qquad
	X_i &\ \stackrel{\text{ind}}{\sim}\  \text{NB}\left(r,\ p_i=\frac{r}{r+E_i R_0}  \right) \label{eq:Xi}
\end{align}
for $i=1,\ldots,K$ and where $r$ and $R_0$ are unknown parameters in the negative binomial model.
Survey-specific rates, $R_i$ have common expectation $\alpha/\beta=R_0$ and standard deviation $\sqrt{\alpha}/\beta=R_0r^{-1/2}$.  The case $r=\infty$ corresponds to no survey-to-survey variation in the underlying rates; i.e.~$R_i=R_0$.  Event counts, $X_i$, have expectations $r(1-p)/p=E_i R_0$, proportional to exposures.  Because multiple surveys inform on parameters $r$ and $R_0$ (equivalently, $\alpha$ and $\beta$) we are not setting them to the neutral gamma values as in Sections~\ref{sec:negbin} and~\ref{sec:twoGroups}, where a prior represented initial uncertainty on a single underlying rate.  
In fact, to keep the analysis within the scope of widely available statistics software, Sections~\ref{sec:extraPoisson} and~\ref{sec:rateChange} use likelihood-based  frequentist methods---maximum likelihood estimates of $(r, R_0)$ and, similarly, likelihood-based uncertainties.  This section is distinctly frequentist, even though the NB likelihood~(\ref{eq:Xi}) is obtained from~(\ref{eq:Ri}) and~(\ref{eq:XiGivenRi}) by application of Bayes' rule.  A fully Bayesian analysis of this setup requires MCMC.

%

The Likelihood Ratio Test (LRT) \citep[see][Section 3.4]{cameron2013regression} for comparing $r=\infty$ (a single common rate) against $r<\infty$ (varied rates) uses the statistic
\begin{align}\label{eq:Delta}
	\Delta &= 2 \left[ \max_{r,R_0}{\cal L}(r, R_0) - \max_{R_0}  {\cal L}(\infty, R_0) \right]
\end{align}
where  $\cal L$ is the log-likelihood function for the negative binomial 
\begin{align}\label{eq:NBPDF}
	{\cal L}(r,R_0) =  \sum_i & \left[ \ln \left(\frac{\Gamma(r+X_i)}{\Gamma(X_i+1)\Gamma(r)}\right) 
	+ r\ln \left(\frac{r}{r+E_i R_0}\right) \right. \nonumber \\
	& \left. +X_i\ln\left(\frac{E_iR_0}{r+E_i R_0}\right)\right].
\end{align}
The first term in $\Delta$ (Equation~\ref{eq:Delta}) requires 2D numerical maximization which could occur at $r=\infty$ indicating no evidence of heterogeneous rates, i.e.~$\Delta=0$.  The second term corresponds to the case of independent Poisson counts with well-known maximum likelihood estimate $\hat R_0\equiv\sum_i X_i / \sum_i E_i$ and maximized log-likelihood equal to 
\begin{align}
   \max_{R_0}&{\cal L}(\infty, R_0) \nonumber \\
   &= \sum_i \left[ X_i \ln (E_i \hat R_0)  -  \ln\Gamma(X_i+1) - E_i \hat R_0 \right]. 
\end{align}

If $\Delta$ exceeds zero, the p-value for homogeneity is 
\begin{align}\label{eq:ODConf}
	\text{p-value} &= {\textstyle\frac{1}{2}}\Pr[\chi^2_1 > \Delta]  
	= \Phi(-\Delta^{1/2})
\end{align}
where the factor ${\textstyle\frac{1}{2}}$ adjusts the usual LRT confidence for a nonstandard limiting distribution of $\Delta$ \citep{lawless1987negative}, $\chi^2_1$ represents a chi-squared random variable with one degree of freedom, and $\Phi(\cdot)$ is the standard normal cumulative distribution function.  A small p-value (e.g.~0.05 or less) is evidence of heterogeneous rates.  With small counts, computed p-values tend to understate the actual evidence \citep{lawless1987negative}.  Monte Carlo simulations can be used to recalibrate p-values if this is a concern.


\subsection{Is a Common Rate Plausible in the Surveys of Table~\ref{tab:data}?}

\begin{figure}[tbp]
\begin{center}	
	\includegraphics[width=\columnwidth]{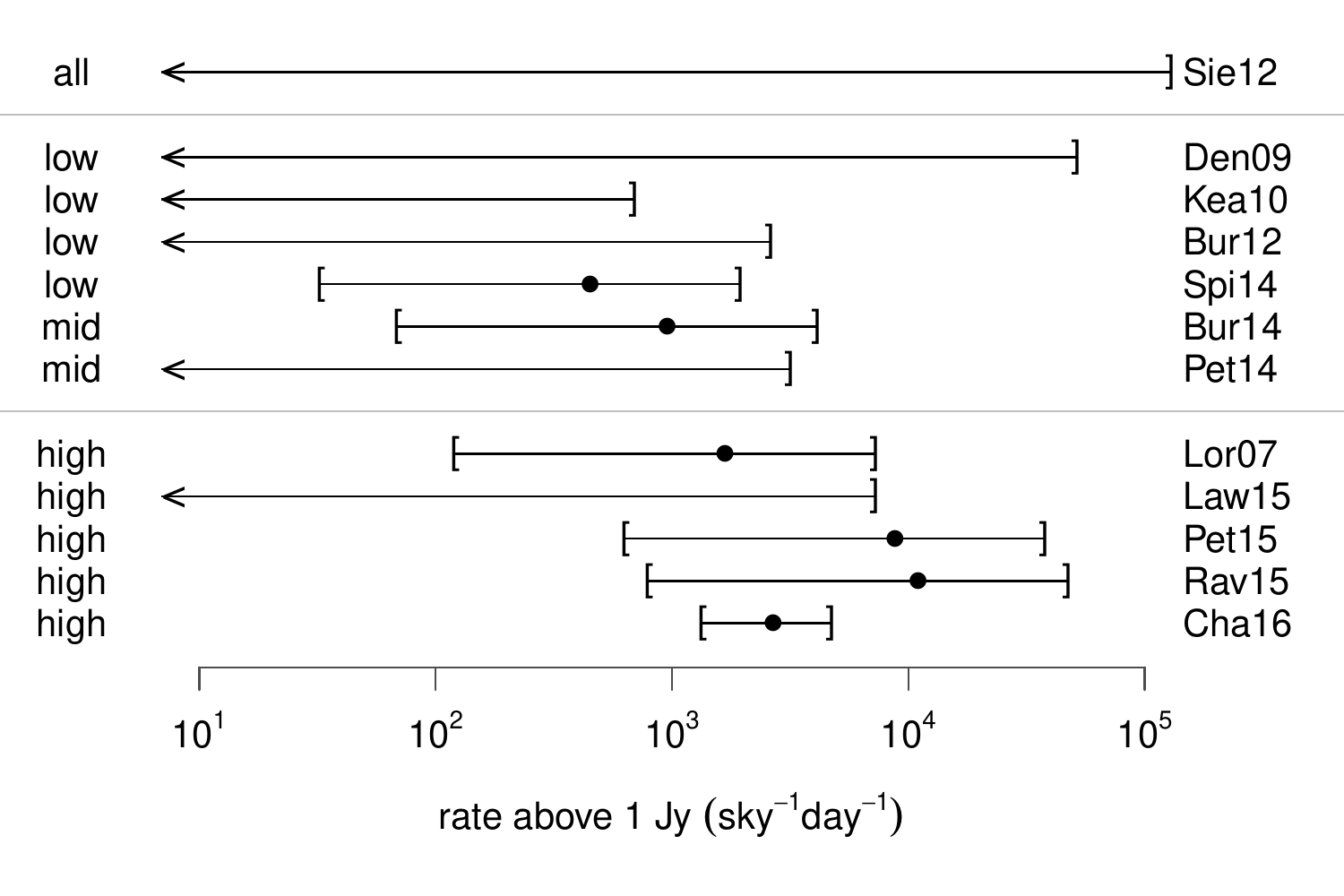}
	\caption{Estimated rates of FRBs above at 1 Jy for each of twelve surveys listed in Table~\ref{tab:data}, assuming Euclidian scaling. The intervals are plotted at the 2.5\% and 97.5\% quantiles of gamma distributions given by Equation~(\ref{eq:RgivenX}).  Points indicate posterior medians for surveys with detections.  Note that the non-detection of Kea10 is discordant with lower bounds of Pet15, Rav15 and Cha16.}
	\label{fig:rates1Jy}
\end{center}
\end{figure}

As in Section~\ref{sec:anisotropic} we standardize exposures to a sensitivity of 1~Jy, assuming Euclidian scaling.  That is, the $(X_i,E_i)$ of Equation~\ref{eq:Xi} are identified with columns $(N_\text{FRB}, E_\text{1Jy})$ in Table~\ref{tab:data}.  Figure~\ref{fig:rates1Jy} plots 95\% credible intervals for FRB rates above 1~Jy.  Endpoints are 2.5\% and 97.5\% quantiles of survey-specific gamma distributions given by Equation~(\ref{eq:RgivenX}).   Median values are shown as points for the surveys that detected one or more FRBs.  Surveys with no detections are shown with left-facing arrows instead of lower bounds.   The figure suggests that a common rate cannot adequately represent these data.  For example, the upper limit of Kea10 is below the lower limits of Cha16, Rav15, and almost Pet15.

The formal likelihood ratio test for extra-Poisson variation proceeds as follows.
Maximized likelihood values are  
$\max_{r,R_0}{\cal L}(r, R_0)= -15.879 $ and 
$\max_{R_0}  {\cal L}(\infty, R_0) = -18.414$.
The corresponding maximum likelihood rates are $\hat R_0 = 1224$ and 1194 sky$^{-1}$ day$^{-1}$, respectively.  
Equation~(\ref{eq:Delta}) gives $\Delta= 2\cdot(18.414 - 15.879)=5.07$
and Equation~(\ref{eq:ODConf}) provides the p-value of $\Phi(-\sqrt{5.07})= 0.012$.
That is, the survey results strongly indicate heterogeneous underlying Poisson rates, assuming Euclidian scaling.  

Heterogeneity is no surprise given the conclusion in  Section~\ref{sec:anisotropic} of a higher rate at high Galactic latitude.  In fact, Figure~\ref{fig:rates1Jy} groups the twelve surveys by Galactic latitude to visually highlight differences.  Applying the likelihood ratio test to only the high latitude data gives $\Delta=0$, meaning no evidence of extra-Poisson variation.  The same holds for the combined mid- and low-latitude subset.  The next section further considers evidence of a higher rate at high latitude.

\section{What is the trend of FRB rate with flux limit?}\label{sec:rateChange}

This section describes and demonstrates fitting of a power-law to the rate, ${\cal R}(F)$, of FRBs exceeding flux $F$:
\begin{align}\label{eq:rateFun}
  {\cal R}(F) \ =\ {\cal R}_0F^b \qquad \Leftrightarrow \qquad \ln {\cal R}(F) = a + b\ln F 
\end{align}
where $a=\ln {\cal R}_0$.  Nominally, one would anticipate that a survey with flux sensitivity limit $F$ would detect a Poisson number of FRBs with expectation equal to ${\cal R}(F)$ multiplied by the time$\times$area exposure of the survey.  Poisson regression would be a good starting point for estimating ${\cal R}(F)$ but an important extension allows for extra-Poisson variation.  \citet{cameron2013regression} discuss relevant  methods and the core ingredient, as might be expected, is the negative binomial distribution.  We now present the basics of NB regression in the context of estimating ${\cal R}(F)$.

\subsection{Negative Binomial Regression}\label{sec:NBRegression}
As previously, let $i=1,\ldots, K$ index a collection of surveys and define
\begin{subequations}
\begin{align}
        F_i &= \text{minimum detectable flux [Jy]} \\
        R_i & =\text{unknown event rate [deg$^{-2}$hr$^{-1}$]  (assuming $S_i=1$)}
\end{align}
\end{subequations}
in addition to the notation of Equation~(\ref{eq:notation}).
We want the expected value of $R_i$ to be given by the power-law curve ${\cal R}(F_i)$ and we also want to allow for random survey biases that are not already included in the known sensitivity factors $S_i$.  Such biases may stem from instrumentation or astronomical factors that are not yet understood.  The model for FRB counts is then (for $i=1,\ldots,K$)
\begin{align} 
        [X_i |R_i] &\ \stackrel{\text{ind}}{\sim}\  \text{Poisson($R_i E_i$)}, 
        \label{eq:NBRegression1} \\
        R_i &\ \stackrel{\text{ind}}{\sim}\  \text{Gamma}\left(\alpha=r, \beta_i=r/{\cal R}(F_i) \right),
        \label{eq:NBRegression2} \\
	\Rightarrow \qquad 
	X_i &\ \stackrel{\text{ind}}{\sim}\  \text{NB}\left(r, p_i=\frac{r}{r+E_i {\cal R}(F_i)}  \right).
	\label{eq:NBRegression}
\end{align}
Survey $i$ has underlying rate, $R_i$ with expectation $\alpha/\beta_i = {\cal R}(F_i)$ and proportional standard deviation $\sqrt{\alpha}/\beta_i = {\cal R}(F_i) r^{-1/2}$. As $r \rightarrow\infty$, the random survey biases become negligible, $R_i={\cal R}(F_i)$, and the model collapses to Poisson regression. For finite $r$ the expected count for survey $i$ is $r(1-p)/p=E_i{\cal R}(F_i)$, the exposure-scaled rate.  As in Section~\ref{sec:extraPoisson}, multiple surveys inform on the parameters in Equation~(\ref{eq:NBRegression2}) and we use the maximum likelihood paradigm for estimation.

\begin{figure*}[t]
\begin{center}	
	\includegraphics[width=0.91\textwidth]{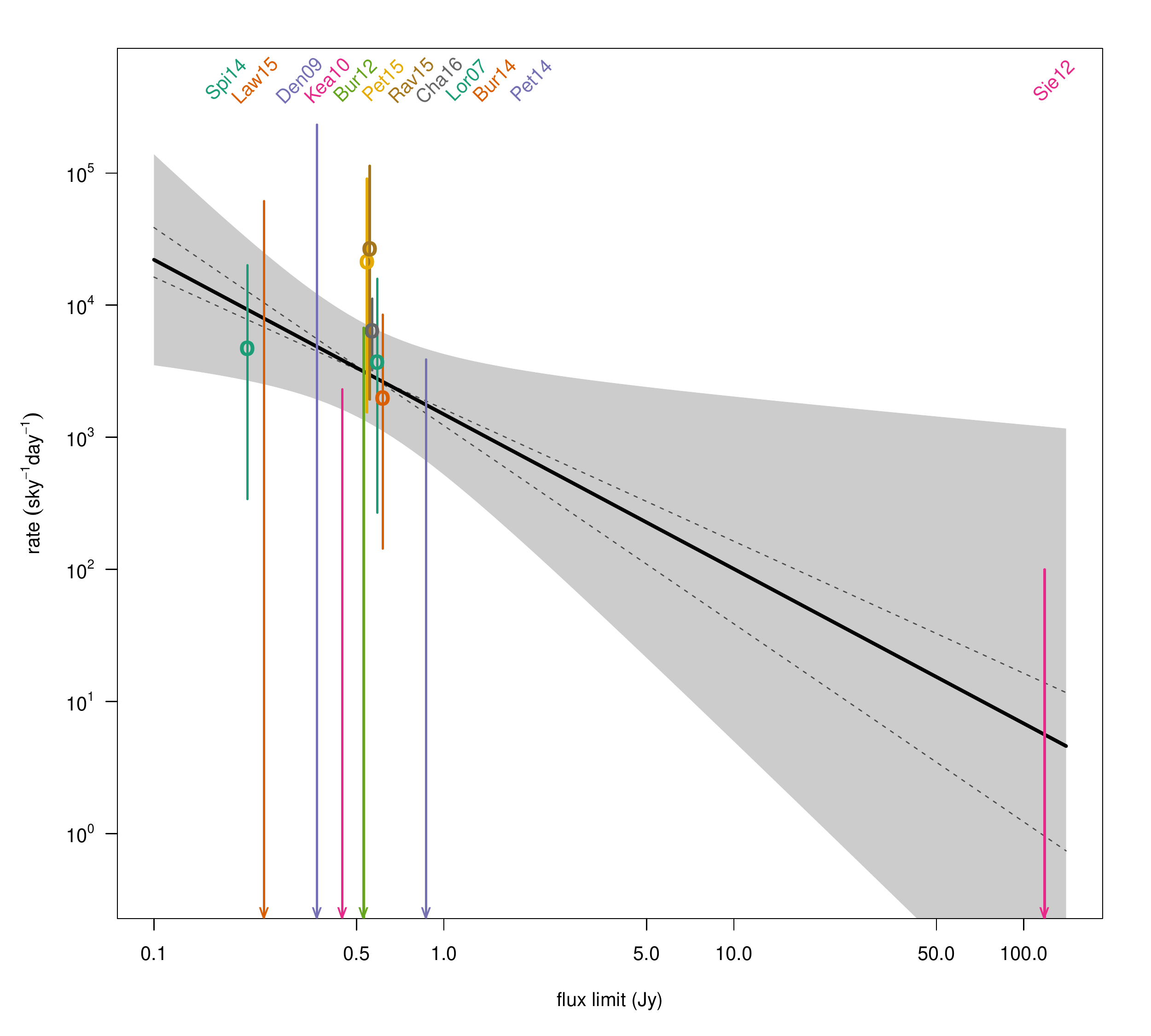}
	\caption{Estimated FRB rate versus flux (solid black) using negative binomial regression on the data in Table~\ref{tab:data}, less Tho13.   The 95\% confidence region for ${\cal R}(F)$ (shaded) includes both Euclidian scaling (steeper dashed, slope $=-3/2$) and b=1 scaling (shallower dashed). A vertical segment for each survey shows the central 95\% credible interval from Equation~(\ref{eq:RgivenX}) using the neutral prior and points show posterior medians for surveys that detected FRBs.}
	\label{fig:rate}
\end{center}
\end{figure*}

Maximum likelihood estimates (and log-likelihood-based  uncertainties) of the parameters $a, b, r$ in Model~(\ref{eq:rateFun}, \ref{eq:NBRegression}) can be obtained from standard GLM software, such as provided in the MASS~\citep[Section 7.4]{venables2002modern} package in R~\citep{Rmanual}.  The calculations below use MASS functions to estimate ${\cal R}(F)$ from twelve FRB surveys.  More generally, the log-linear form of ${\cal R}(F)$ in Equation~\ref{eq:rateFun} is readily generalized to include other additive predictors beyond $\ln F$ and we use this capability below to fit different intercepts to low/mid- vs.~high Galactic-latitude surveys.  Online supplemental materials provide simple R scripts showing how to use the software.


Figure~\ref{fig:rate} plots rate estimates and uncertainties from the data in Table~\ref{tab:data} (except Tho13). 
Points are medians of  gamma distributions given in Equation~(\ref{eq:RgivenX}) with the neutral prior ($\alpha=1/3, \beta=0$); vertical segments are 95\% credible intervals.  Points are omitted for surveys with no detections.  The solid trend line is the maximum likelihood estimate of ${\cal R}(F)$ using negative binomial regression to fit the three parameters ($a,b,r$) in Equations~(\ref{eq:rateFun}) and~(\ref{eq:NBRegression}).  The (negative) slope ($\hat b=1.17$) lies between the dashed reference slopes $b=3/2$ (Euclidian) and $b=1$ (inverse flux).  The shaded region shows 95\% confidence bands for ${\cal R}(F)$.  From the power-law fit, the estimated rate of FRBs with flux above 1~Jy is $\hat{\cal R}(1\ \text{Jy})=1491$ $\text{sky}^{-1}\text{day}^{-1}$ with 95\% confidence interval (521, 4268).  The estimate from Euclidian scaling (steeper dashed) is 1224 $\text{sky}^{-1}\text{day}^{-1}$ above 1~Jy, an exact match to the value computed in Section~\ref{sec:extraPoisson}.  

\begin{figure*}[t]
\begin{center}	
	\includegraphics[width=0.91\textwidth]{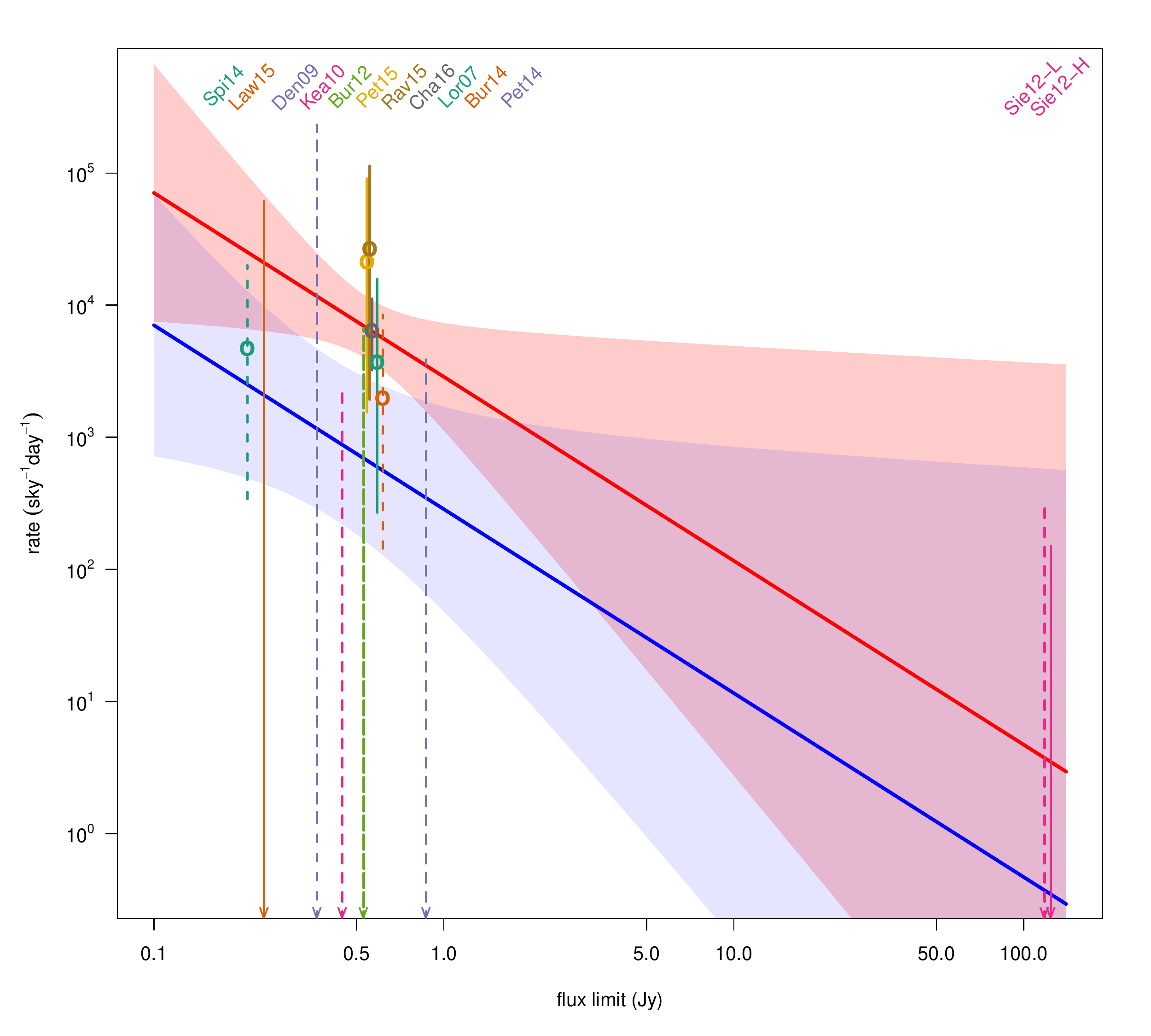}
	\caption{Split estimates of FRB rates vs.~flux limit for high-Galactic-latitude surveys (solid intervals and red trend line) and mid/low latitudes (dashed intervals and blue trend line).  The common estimated slope is $-\hat b = -1.39$.  Shaded regions indicate uncertainty on the trends.  The difference in trend lines is statistically significant; see Table~\ref{tab:models}. Total exposure for Sie12 is allocated 1/3 to low/mid Galactic latitudes and 2/3 to high latitudes. }
	\label{fig:rateHighLow}
\end{center}
\end{figure*}

All vertical segments for the individual surveys  intersect the confidence bands but a single trend line cannot be drawn through all of them, so there is some visual evidence of extra-Poisson variation.  In particular, Kea10 (low latitude) and Cha16 (high latitude) are discordant.
A likelihood ratio test for extra-Poisson variation can be conducted by extending the definition of $\Delta$ (Equation~\ref{eq:Delta}) in Section~\ref{sec:extraPoisson} to include $b$ in the maximizations.  The computed value is $\Delta=3.99$, which is fairly large as indicated by a small p-value of 0.023 (Equation~\ref{eq:ODConf}); i.e., the data do not agree with an assumption of homogeneous power-law rates across the surveys ($R_i={\cal R}(F_i)$).  Equation~(\ref{eq:NBRegression2}) implies $[R_i/{\cal R}(F_i)]\sim\text{Gamma}(r,r)$, with standard deviation $r^{-1/2}$.  The maximum likelihood estimate of $r$ (Equation~{\ref{eq:NBRegression}}) is $\hat{r}=1.219$ so the underlying rates $R_i$ deviate from the all-sky rate by $\hat{r}^{-1/2}=91\%$ (at 1$\sigma$).

The regression model can easily be extended to estimate different rates at high vs.~mid and low Galactic latitudes, with a common source count index, $b$:
\begin{equation}\label{eq:splitRate}
	\ln {\cal R}(F) =
		 \begin{cases} 
		 	a_\text{high} + b\ln F, & \text{for high Galactic latitude,} \\	
			a_\text{med} + b\ln F, & \text{for mid or low latitude.} 
		\end{cases}
\end{equation}

\begin{table*}[tb]
\centering
\begin{tabular}{cc@{\hskip 5ex}cc@{\hskip 0ex}c}
& Figure~\ref{fig:rate} & \multicolumn{2}{l}{\quad------ Figure~\ref{fig:rateHighLow} ------} & \citet{lawrence2016} \\
 & all-sky & low/mid & high & high \\ \hline
 \multicolumn{1}{l}{$\hat{\cal R}$(1 Jy) [{sky}$^{-1}$day$^{-1}$]} & 1491 & 285 & 2866 & 1315\\
\multicolumn{1}{c}{95\% conf.} & (521,4268) & (48, 1701) & (1121, 7328) & (510, 2152) \\ 
\multicolumn{1}{l}{$\hat{\cal R}$(2 Jy ms)}  & 2397 & 502 & 5042 & 1814 \\
\multicolumn{1}{c}{95\% conf.} & (998, 5754) & (110, 2279) & (2775, 9162) & (1030, 3196) \\ 
\multicolumn{1}{l}{$\hat{\cal R}$(4 Jy ms)} & 1065 & 191 & 1920 & 1047 \\
\multicolumn{1}{c}{ 95\% conf.} & (308, 3681) & (25, 1466) & (547, 6744) & (501, 2187)\footnote{This is the reported 99\% confidence interval.} \\ \hline
 $\hat b$, index & 1.17 & \multicolumn{2}{c}{1.39} &  0.79 \\
 95\% conf. & (0.19, 2.15) & \multicolumn{2}{c}{(0.12, 2.67)} & (0.38, 1.22) \\ \hline 
 std.dev. $R_i/{\cal R}(F_i)$ & 91\% & \multicolumn{2}{c}{0\% (fit)} & 0\% (defined) \\ 
 \end{tabular}
\caption{Summary of model fits shown in Figure~\ref{fig:rate} and~\ref{fig:rateHighLow}.  $\hat{\cal R}$(1 Jy) is the estimated rate above 1~Jy flux whereas $\hat{\cal R}$(2 Jy ms) and $\hat{\cal R}$(4 Jy ms) are estimates above stated {\em fluence} limits, assuming a typical 3 ms pulse; $\hat b$ is the power-law source count index.  The final row reports extra-Poisson variation; that is, the estimated standard deviation of survey-specific relative biases in rates.  Whereas 91\% extra-Poisson variation is needed for the all-sky trend, none is needed for the split-latitude trends.  Estimates in the final column, \citet{lawrence2016}, are discussed in Section~\ref{sec:trendComments}.}  
\label{tab:models}
\end{table*}

Figure~\ref{fig:rateHighLow} plots rate estimates.  In this case we split the total exposure of Sie12, assigning 1/3 to the mid/low latitudes and 2/3 to the high latitudes, in rough correspondence to the drift-scan used in that survey.  Surveys categorized as low or medium Galactic latitude are shown with dashed credible intervals, whereas  solid-line intervals indicate high latitude surveys.  The cluster of (Parkes) high latitude surveys with sensitivities at about 0.55 Jy necks-down the shaded 95\% credible bands in that region.  Similar necking in the mid/low latitude bands opens a gap around 0.5 Jy, indicating the high latitude rate is statistically above that at mid/low latitudes.  With this split-latitude trend, the estimated high latitude rate above 1~Jy flux is 2866 [{sky}$^{-1}$day$^{-1}$] with 95\% confidence interval $(1121, 7328)$.  Relative uncertainty of the rate is minimized at ~0.5~Jy where the 95\% interval covers a factor of ~3, even after the vast amount of observing represented by this collection of surveys.  To reduce this uncertainty to, say $\pm 10\%$ (at 95\% confidence) would require a factor of $(3/0.2)^2=225 (!)$ times as much survey exposure.  Constraining the FRB rate to this extent will require large area telescopes like the Canadian Hydrogen Intensity Mapping Experiment \citep{bandura2014chime}.

Table~\ref{tab:models} summarizes differences between model fits, with {\em all-sky} referring to the single-trend fit of Figure~\ref{fig:rate} and the pair of {\em low/mid} and {\em high} referring to the split-latitude trends of Figure~\ref{fig:rateHighLow}.  Rate estimates are given for sensitivity thresholds of 1~Jy flux and 2 and 4 Jy~ms fluence.  Fluence-based rates assume a 3~ms pulse and are provided for comparison with two currently prominent estimates:  \citet{champion2016fivenew} estimate 2500  $\text{sky}^{-1}\text{day}^{-1}$ above ${\sim}$2 Jy ms, about half our estimate of 5402 and below our 95\% confidence interval (2775, 9162).  \citet{rane2016search} estimate 4400 above 4 Jy ms, more than double our estimate of 1920 but we think their method should have produced 3188 sky$^{-1}$ day$^{-1}$ (see discussion below).  Either value falls within our 95\% confidence interval (547, 6744).   The final column of Table~\ref{tab:models} is discussed in Section~\ref{sec:trendComments} below. 

The source count index does not change much between the two model fits relative to its uncertainty, with values of $b=3/2$ and $b=1$ easily within the 95\% confidence intervals.  Note, however, that the low sensitivity (high flux limit) of \citet{siemion2012search} gives that single survey substantial leverage in fitting $b$.  One might wish for an additional low-sensitivity survey to reduce uncertainty but there is a better way.   \citet{lawrence2016} show how to incorporate the detected flux values to into estimation of ${\cal R}(F)$ with no need for surveys covering a range of sensitivities.  

One nice property of the split-latitude fit is that there is essentially no evidence of extra-Poisson variability as shown by the fitted standard deviation of $R_i/{\cal R}(F_i)$ equal to 0\% in the last row of Table~\ref{tab:models}.  That is, the two trend lines fit to all twelve surveys with no statistical evidence of survey-specific biases.   

Our final comments concern the FRB rate estimate of \citet{rane2016search} who use Bayes' formula (their Equation~(4)) with a uniform prior distribution, the special-case gamma distribution with $\alpha=1, \beta=0$.  This should produce slightly higher rate estimates that our neutral recommendation ($\alpha=1/3, \beta=0$).  However, we cannot reproduce basic calculations in \citet{rane2016search}.  For example, the form of posterior given in Rane's Equation~(6) is correct, and represents an exponential distribution with mean value $K_1^{-1}=(f T_P A_P)^{-1}=1484$ sky$^{-1}$ day$^{-1}$, but Rane~et~al. calculate 220.  Using data from Rane's Table~2 and their uniform prior we calculate an FRB rate of 3188 sky$^{-1}$ day$^{-1}$ above 4 Jy~ms, quite different from Rane's value of 4400.   Using the neutral gamma prior with Rane's data gives an even lower estimate of 2922 sky$^{-1}$ day$^{-1}$ above 4 Jy~ms, in substantial better alignment with our 4 Jy~ms estimate of 1920 at high Galactic latitude.

\subsection{Comparison to \citet{lawrence2016} and Other Considerations}\label{sec:trendComments}

The above negative binomial regression analysis follows the common practice of pairing half-max sensitivity with  beam area at half-max as if the beam had a top-hat form with area and sensitivity given by FWHM values.  \citet{lawrence2016} utilize a non-homogeneous Poisson process model that incorporates the actual beam form and fits to the collection of measured flux values for all FRBs detected in the surveys.  Interestingly, a FWHM top-hat closely approximates the effective area at half-max sensitivity if the beam is Gaussian and the source count index is $b=3/2$, Euclidian scaling, but understates the effective area for smaller values of the index.

The final column of Table~\ref{tab:models} gives high Galactic latitude estimates from \citet{lawrence2016}.  Our rate above 1~Jy at high latitude is more than double theirs and their source count index is only $\hat b=0.79$, compared to our 1.39.   Their statistical uncertainty on $b$ is also much less than ours.  On the surface the analyses seem inconsistent.  However,  \citet{lawrence2016} also demonstrate that a source count index of $b=0.79$ implies  an equivalent beam area that is 83\% ($=(0.79\ln 2)^{-1}-1$) {\em larger} than the top hat approximation used here.  The value $b=0.79$ is well within our uncertainty and adjusting our rate for the larger effective area gives an adjusted rate of 1569 (=2866/1.83) sky$^{-1}$ day$^{-1}$, only 19\% higher than the \citet{lawrence2016} estimate.  So, differences in the analyses are understandable in terms of the relation between effective beam area and source count index.  This comparison demonstrates the importance of FRB researchers gaining confidence as to whether the true index is substantively below the Euclidian value of 3/2.


The regression approach illustrated above can easily be extended to estimate the trend in rate as a function of flux limit and additional predictors such as a continuous-valued Galactic latitude or sky temperature. We caution, however, that searching across many explanatory predictors is likely to `discover' an apparent trend that will not be statistically well-substantiated with so few surveys.  

Finally, the NB model entertains a different random bias for each survey in the collection and, therefore, if multiple surveys are expected to have a common bias they should be defined as a single survey, with an aggregate exposure and event count, before applying the negative binomial regression.  Our handling of Sie12 in the split analysis breaks from this advice but with no evidence of extra-Poisson variation the issue is moot.  Similarly, if portions of surveys potentially have different biases, these should be disaggregated before the regression analysis.

\section{Summary and Discussion}\label{sec:summary}
While rare event counts follow Poisson probabilities, best practices for making statistical conclusions with count data usually require going beyond straight Poisson calculations.  In particular, if uncertainty or variation in underlying Poisson rates is represented by a gamma distribution, then the rare event counts are distributed as negative binomial random variables.  We have shown how to use the Poisson-gamma construction to answer a variety of questions pertinent to statistical inference on rare event rates in transient science.  For situations with only a single rate to be estimated, or two rates to compare (Sections~\ref{sec:hours} and~\ref{sec:twoGroups}) we advocate following the Bayesian paradigm with a so-called neutral gamma prior distribution \citep{kerman2011neutral}.  The paradigm produces direct answers to questions about probabilities of detecting events given data from past surveys, avoiding 
overly-conservative answers associated with frequentist methods.  

Meta-analyses combine data across a number of surveys and should entertain the possibility of heterogeneity in underlying rates  from survey differences that are not yet accounted for through known sensitivity factors.  Sections~\ref{sec:extraPoisson} and~\ref{sec:rateChange} explained  (non-Bayesian) methods for analyzing data from a collection of surveys where extra-Poisson variation is a distinct possibility.  Sections~\ref{sec:extraPoisson} and~\ref{sec:rateChange} illustrated the use of these methods on data from twelve FRB surveys.

This paper is primarily a tutorial on statistical methods for working with rare event count data.  Some essential points are: 1) rare event rates are never known exactly and with small numbers of detections, rate uncertainty is an important aspect of the analysis;  2) Poisson and negative binomial regressions are core methods in the arena of Generalized linear models.  GLM software makes it straightforward to fit trends in  event rates with respect to multiple predictor variables such as  flux threshold, Galactic latitude, and sky temperature,  and to account for extra variation across surveys beyond the base Poisson statistics.  Uncertainties on rates and on trend coefficients will be incorrect if Poisson statistics are assumed  when known sensitivity factors do not completely account for heterogeneity of rates.

Some of the results exemplified in this paper are scientifically interesting in their own right.  Namely, we find 
1) modest evidence ($p=0.07$) from two surveys  \citep{ravi2015fast, petroff2015survey} that longer observations on fewer fields is more productive than the `scan wide' style of HTRU \cite[(Section~\ref{sec:wideDeep}]{champion2016fivenew}; 
2) four low-Galactic-latitude surveys demonstrate a highly significant lower rate ($p=5\times 10^{-5}$) of FRBs relative to five high latitude surveys (Section~\ref{sec:anisotropic}). The mid vs.~high latitude comparison is marginally significant ($p=0.03)$.
3)  the FRB rate at high Galactic latitude is estimated to be 2866 $\text{sky}^{-1}\text{day}^{-1}$ above 1 Jy with 95\% confidence interval (1121, 7328) from a meta-analysis of all twelve surveys in Table~\ref{tab:data}, allowing for general power-law scaling with flux sensitivity and extra-Poisson variation (Section~\ref{sec:rateChange}).  The corresponding combined low/mid-latitude rate is  285 $\text{sky}^{-1}\text{day}^{-1}$ above 1 Jy with 95\% confidence interval (48, 1701), lower than the high latitude rate with high statistical significance ($p=0.003$).  Our high latitude results are in rough agreement with other recent estimates.  \citet{champion2016fivenew} estimate 2500  $\text{sky}^{-1}\text{day}^{-1}$ above a {\em fluence} of ${\sim}$2 Jy ms, about half our estimate of 5042 assuming a 3~ms pulse.  On the other hand,  \citet{rane2016search} estimate 4400 above 4 Jy ms, more than double our estimate of 1920.  See Table~\ref{tab:models} for confidence intervals. We could not reproduce the value of 4400 in \citet{rane2016search}.  Our calculation based on their input data and analysis method gives 3188 $\text{sky}^{-1}\text{day}^{-1}$, in better alignment with the analysis presented here. 

The suggestion that longer observations are more productive is odd.  One model that could explain the apparent difference in rates is as follows.  Suppose all FRB sources generate repeat pulses at rates that vary source-to-source and suppose there are few enough sources that a random pointing has a significant chance of containing no detectable source.  This finite source count model of repeaters does not follow the key assumption of a Poisson process as stated in the Introduction and the apparently higher `look long' rate relative to the `scan wide' rate could arise from either good luck in choice of a small number of pointings or possibly from publication bias in which detections are more likely to be reported than non-detections, especially in smaller scale studies.  

A statistical model could be constructed to reflect a finite number of sources with random repeat rates.  Gamma distributions would represent uncertainties in separate spatial and temporal rates and these would be combined with nested Poisson distributions for the numbers of FRB sources and detections associated with each pointing in a survey.   Analysis of this setup would require complete pointing lists and we would recommend Monte Carlo Markov Chain (MCMC) fitting of a Bayesian model.  The analysis could jointly bound spatial and temporal rates associated with FRB sources but uncertainties would be large.  Although MCMC is beyond the scope of this paper the above outline is suggestive of the flexibility available within the Bayesian paradigm. See references \citep{gelman2014bayesian, carpenter2015stan} for general Bayesian modeling methods and tools.

\section{Acknowledgements}
This work was supported by the University of California Lab Fees Program under award number \#LF-12-237863.

\appendix

\section{Frequentist method for hours required to observe a pulse}\label{sec:frequentist}
The question of how long to observe can also be addressed with frequentist prediction bounds.  The frequentist approach is more common and removes the need to specify an appropriate prior distribution for $R$ but comes at the expense of overly conservative answers and convoluted interpretations relative to the Bayesian paradigm.     

The frequentist approach regards $R$ as having a fixed but unknown value; it has no prior distribution.
With past count $X_1$, a lower 95\% prediction bound on the future count $X_2$ is a value $L\equiv L(X_1;E_1,E_2)$ that $X_2$ will meet or exceed with probability at least 0.95.  That is,
\begin{align}\label{eq:prediction}
	\Pr \left[X_2 \geq L(X_1; E_1,E_2) \right] \ \geq \ 0.95
\end{align}

This probability is calculated with respect to the independent randomness of {\em both} $X_1$ and $X_2$ (Equation~\ref{eq:xi}) and the condition must hold for every assumed value of $R>0$.  With the first count in hand we compute a number $L\equiv L(X_1; E_1,E_2\text{=}T_2 A_2 S_2)$, and make the standard interpretive statement that at least $L$ pulses are expected in $T_2$ additional hours with 95\% predictive confidence.  The interpretation  is awkward because condition~(\ref{eq:prediction}) refers to joint randomness of $X_1,X_2$ whereas the confidence statement is constructed from a {\em given} value of $X_1$.  But this is the standard frequentist paradigm for a prediction bound.

The most common lower 95\% prediction bound for a future Poisson count \citep{hahn2011statistical, nelson1970confidence} is the largest integer $L$ such that
\begin{equation}\label{eq:countBound}
	 \frac{X_1}{E_1} \ \leq\  \frac{(L+1)}{E_2} \cdot Q_F(0.95; m=2L+2, n=2X_1)
\end{equation}
where  $Q_F(0.95; m,n)$ is the 0.95 quantile of the $F$ distribution with $m$ and $n$ degrees of freedom. 
Use of an $F$ distribution in this context is not obvious; \citet{nelson1970confidence} derives the method in an appendix, achieving~(\ref{eq:prediction}) for every conditional distribution $[X_2\ |\ X_1+X_2]$ and thereby conservatively achieving it with respect to the joint distribution $[X_1,X_2]$.

Setting $L=1$ in~(\ref{eq:countBound}), substituting~(\ref{eq:exposure}), and solving for $T_2$ produces the number of hours required to obtain 95\% predictive confidence of detecting at least one pulse.  For the example problem, $X_1=4, T_1=551.9, A_1 S_1=A_2 S_2$ and we obtain
\begin{equation}
	T_2 = \frac{1+1}{4/551.9} \cdot Q_F(0.95; 4,8) = \frac{1103.8}{4} \cdot 3.838 = 1059 \text{ hours}.
\end{equation}	
This is {\em nearly double} the $550$ hours required by the Bayesian solution.

Which method is correct?
Perhaps surprisingly, the statistics community is not settled on best confidence bounds to use for discrete distributions such as Poisson or Binomial counts.  \citet{agresti1998approximate} argue that frequentist bounds are too conservative, with  coverage probabilities (e.g., the left side of~(\ref{eq:prediction})) that are too far above their nominal levels (e.g., 0.95) for routine practical use. \citet{krishnamoorthy2011improved} review various options  to reduce conservatism (see also \cite{agresti2001small}) but there is no generally accepted best solution; practitioners must weigh factors such as simplicity, interpretability, and the importance of a conservative confidence level verses one that is correct when averaged over a range of possible rates, $R$.

We advocate the Bayesian paradigm of Section~\ref{sec:negbin} for two reasons.  First, ease of interpretation: given $X_1$, the probability of observing $X_2>0$ is computed directly.  This contrasts with the awkward interpretation of confidence in the frequentist paradigm.  Second, Bayesian bounds do not suffer from built-in poor calibration.  Exact frequentist bounds require the {\em minimum} coverage to be no less than the nominal value (i.e., Equation~(\ref{eq:prediction}) holds for {\em every} possible $R$) leading to overly-conservative bounds, whereas Bayesian bounds exactly meet the coverage requirement {\em on average} over the posterior distribution of $R$.  Criticism of the Bayesian paradigm revolves around the difficulty of specifying an appropriate prior distribution for the unknown rate parameter.  \citet{kerman2011neutral} make a convincing case to use the neutral gamma prior ($\alpha=1/3, \beta=0$) as a generic choice, especially appropriate for small counts when little prior information is available about the rate.  One slight caution is that for a count of zero, the posterior distribution, Equation~(\ref{eq:RgivenX}), has a lower 0.05 quantile of ${\sim}10^{-4}E_1^{-1}$, implying that the exposure $E_1$ is large enough to produce a minimum of ${\sim}10^{-4}$ chance of detection.  This is typically a modest assumption.

\bibliography{howLong}
\bibliographystyle{aasjournal}

\end{document}